\documentclass[letterpaper,11pt]{article}

\usepackage[left=1in,right=1in,top=1in,bottom=1in]{geometry}
\usepackage{amsmath}
\usepackage{amssymb}
\usepackage{amsthm}
\usepackage{hyperref}
\usepackage{times}
\usepackage{xcolor}

\theoremstyle{plain}
\newtheorem{theorem}{Theorem}[section]

\newtheorem{lemma}[theorem]{Lemma}

\theoremstyle{definition}
\newtheorem{defn}[theorem]{Definition}
\newtheorem{example}[theorem]{Example}

\DeclareMathOperator*{\E}{\mathbb{E}}
\DeclareMathOperator*{\Var}{Var}
\DeclareMathOperator*{\Cov}{Cov}
\DeclareMathOperator{\Stab}{\mathbb{S}}

\DeclareMathOperator{\sgn}{sgn}
\newcommand{\scalprod}[1]{\left<  #1 \right> }
\newcommand{\R}{\mathbb{R}}

\newsavebox{\algbox}
\newenvironment{alg}[1]{%
  \providecommand{\alginput}{}
  \providecommand{\algoutput}{}
  \providecommand{\algtitle}{}
  \renewcommand{\alginput}{\item[] \textsc{Input:} }
  \renewcommand{\algoutput}{\item[] \textsc{Output:} }
  \renewcommand{\algtitle}{#1}
  \begin{figure}[hbt]
  \center
  \begin{lrbox}{\algbox}
  \begin{minipage}[c]{0.95\textwidth}
  \begin{enumerate}
}{%
  \end{enumerate}
  \end{minipage}
  \end{lrbox}
  \framebox{\usebox{\algbox}}
  \caption{\algtitle}
  \label{alg:\algtitle}
  \end{figure}
  \renewcommand{\alginput}{\algerror}
  \renewcommand{\algoutput}{\algerror}
  \renewcommand{\algtitle}{\algerror}
}

\begin{document}
\begin{titlepage}
\thispagestyle{empty}
\title{On the Usefulness of Predicates}
\author{
  Per Austrin
  \thanks{Funded by NSERC.}\\
  University of Toronto 
  \and 
  Johan H{\aa}stad
  \thanks{Funded by ERC Advanced Investigator Grant 226203.}\\
  KTH Royal Institute of Technology
}

\maketitle

\begin{abstract}
Motivated by the pervasiveness of strong inapproximability results for
Max-CSPs, we introduce a relaxed notion of an 
approximate solution of a Max-CSP.  In this
relaxed version, loosely speaking, the algorithm is allowed to replace
the constraints of an instance by some other (possibly real-valued)
constraints, and then only needs to satisfy as many of the new
constraints as possible. 

To be more precise, we introduce the following
notion of a predicate $P$ being
\emph{useful} for a (real-valued) objective $Q$: given an almost
satisfiable Max-$P$ instance, there is an algorithm that beats a
random assignment on the corresponding Max-$Q$ instance applied
to the same sets of literals.  The standard notion of 
a nontrivial approximation algorithm for a Max-CSP with
predicate $P$ is exactly the same as saying that $P$ is
useful for $P$ itself.

We say that $P$ is useless if it is not useful for any $Q$.  This
turns out to be equivalent to the following pseudo-randomness
property: given an almost satisfiable instance of Max-$P$ it is hard
to find an assignment such that the induced distribution on 
$k$-bit strings defined by the instance is not essentially uniform.

Under the
Unique Games Conjecture, we give a complete and simple
characterization of useful Max-CSPs defined by a predicate: such a
Max-CSP is useless if and only if there is a pairwise independent
distribution supported on the satisfying assignments of the predicate.
It is natural to also consider the case when no negations are allowed
in the CSP instance, and we derive a similar complete characterization
(under the UGC) there as well.

Finally, we also include some results and examples shedding additional
light on the approximability of certain Max-CSPs.

% We prove this both in the case when negated literals are allowed and in the case when negations are not allowed.
\end{abstract}
\end{titlepage}
 
\section{Introduction}

The motivation for this paper comes from
the study of maximum constraint satisfaction
problems (Max-CSPs).  We are given a sequence of
constraints, each depending on a constant number
of variables, and the goal is to find an assignment 
that maximizes the number of satisfied constraints.
Essentially any such problem is NP-hard and
a large number of papers have studied the
question of approximability of this class of problems%
%\cite{jhacm,kstw,ow:09,xyz}
.
The standard concept of approximability is that an algorithm is a
$C$-approximation algorithm if, on any instance $I$, it outputs a
number $A(I)$ such that $C \cdot O(I) \le A(I) \le O(I)$, where $O(I)$
is the optimum value of $I$.

There are finer measures of performance.  For example, one can take
$C$ above to be a function of the optimum value $O(I)$.  That is, for
each fraction of constraints satisfied by the optimal solution, we try
to determine the best solution that can be found efficiently.  The
only problem where this has been fully done explicitly is Max-Cut,
where, assuming the unique games conjecture, O'Donnell and Wu
\cite{ow:08} has found the entire curve of approximability.  In a
remarkable paper, Raghavendra \cite{raghavendra} showed that, assuming
the unique games conjecture, the best such approximation possible is
the one given by a certain natural semidefinite programming-based
(SDP) relaxation of the problem.  However, understanding the
performance of this SDP is difficult in general and in this paper we
are interested in more explicit bounds.

Max-Cut turns out to be approximable in a very
strong sense.  To describe these results note
that for Max-Cut it is the case that a random
assignment satisfies half of the constraints
on average.  Whenever the optimum satisfies a
fraction $1/2 + \epsilon$ of the constraints then
it is possible to efficiently find an assignment
that satisfies a fraction $1/2 + c(\epsilon)$ of the constraints
where $c(\epsilon)$ is strictly
positive, depending on $\epsilon$ \cite{cw}.  In other words, whenever
the optimal solution satisfies a non-trivial fraction
of the constraints it is possible to efficiently find an assignment
that satisfies a smaller, but still non-trivial, fraction
of the constraints.

In this paper the main focus is on the other end of the spectrum.
Specifically we are interested in the following property: even if
the optimal solution satisfies (almost) all the constraints it is
still hard to find an assignment that satisfies a non-trivial
fraction.  This might sound like an unusual property, but evidence is
mounting that most CSPs have this property.  We say that such a
CSP is \emph{approximation resistant} (a formal definition appears in Section~\ref{sec:prelim}).

We shall focus on a special class of CSP defined by a single predicate
$P: \{-1,1\}^k \rightarrow \{0,1\}$ (throughout the paper we identify
the Boolean value true with $-1$ and false with $1$).  Each constraint
asserts that $P$ applied to some $k$ literals (each literal being
either a variable or a negated variable) is true.  We refer to this
problem as Max-$P$, and say that $P$ is approximation resistant if
Max-$P$ is.

Several predicates are proven to be approximation resistant in
\cite{jhacm} and the most notable cases are when the predicate in
question is the XOR, or the usual OR, of 3 literals.  For the
latter case, Max-3Sat, it is even the case that the hardness remains
the same for satisfiable instances.  This is clearly not the case for
XOR since a satisfying assignment, if one exists, can be
found by Gaussian elimination.  Hast \cite{hast} studied predicates of
arity 4 and of the (exactly) 400 different predicates, 79 are proven
to be approximation resistant, 275 are found to be non-trivially
approximable while the status of the remaining 46 predicates was not
determined.  Some results exist also for larger predicates and we
return to some of these results in Section~\ref{sec:easyuseless}.  If
one is willing to believe the unique games conjecture (UGC) of Khot
\cite{khotunique} then it was established in \cite{ah11} that an
overwhelming majority of all predicates are approximation resistant.
This paper relies on a result \cite{AustrinMossel:08} establishing
that any predicate $P$ such that the set of accepted strings $P^{-1}(1)$ supports a
pairwise independent distribution is, assuming the UGC, approximation resistant.

In spite of all these impressive results we want to argue that
approximation resistance is not the ultimate hardness condition
for a CSP.  Approximation can be viewed as relaxing the requirements:
if there is an assignment that satisfies a large number,
or almost all, of a given set of constraints, we are content
in finding an assignment that satisfies a lower but still
non-trivial number of constraints.  In some situations,
instead of relaxing the number of constraints we want to satisfy 
it might make more sense to relax the constraints themselves.

Sometimes such relaxations are very
natural, for instance if considering a threshold predicate we
might want to lower the threshold in question.
It also makes sense to have a real-valued measure
of success.  If we give the full reward for satisfying the
original predicate we can have a decreasing
reward depending on the distance to the closest
satisfying assignment.  This is clearly natural in
the threshold predicate scenario but can also
make good sense for other predicates.

It seems like the least we can ask for of a CSP is that
when we are given an instance where we can satisfy
(almost) all constraints under a predicate $P$
then we can find an assignment that does something
non-trivial for some, possibly real-valued, relaxation $Q$.
This brings us to the key definition of our paper.  

\begin{defn}
The predicate $P$ is {\em useful} for the real-valued function $Q: \{
-1, 1\}^k \rightarrow \R$, if and only if there is an $\epsilon>0$ such that
given an instance of Max-$P$ where the optimal solution satisfies a
fraction $1-\epsilon$ of the constraints, there is a polynomial time
algorithm to find an assignment $x^{0}$ such that
$$
\frac 1m \sum_{j=1}^m Q(\bar x^{0}_j) \geq \E_{x \in \{-1,1\}^k}[Q(x)]+\epsilon.
$$
Here $\bar x_j^{0}$ denotes $k$-bit string giving the
values of the $k$ literals in the $j$'th constraint
of $P$ under the assignment $x^{0}$. 
\end{defn}

Given a notion of ``useful'' it is natural to define ``useless''.  We
say that $P$ is useless for $Q$ if, assuming P $\ne$ NP, it is not
useful for $Q$.  We choose to build the assumption P $\ne$ NP into the
definition in order not to have to state it for every theorem -- the
assumption is in a sense without loss of generality since if P $=$ NP
then uselessness is the same as a related notion we call
information-theoretic uselessness which we briefly discuss in
Section~\ref{sec:info}.  Note that uselessness is a generalization of
approximation resistance as that notion is the property that $P$ is
useless for $P$ itself.

This observation implies that requiring a predicate to
be useless for any relaxation is a strengthening
of approximation resistance.  The property
of being a relaxation of a given predicate is somewhat in the eye
of the beholder and hence we choose the following definition.
\begin{defn}
The predicate $P$ is {\em (computationally) useless} if
and only if it is useless for every $Q:
\{-1,1\}^k \rightarrow \R$.
\end{defn}
As described in Section~\ref{sec:easyuseless}, it turns out that almost all
approximation resistance proofs have indeed
established uselessness.   There is a natural reason that we get this
stronger property.
In a standard approximation resistance
proof we design a probabilistically checkable proof
(PCP) system where the acceptance criteria is given
by $P$ and the interesting step in the proof
is to analyze the soundness of this PCP.
In this analysis we use the Fourier expansion of $P$ and
it is proved that the only term that gives a significant
contribution is the constant term.  The fact that
we are looking at the same $P$ that was used to
define the PCP is usually of no importance.
It is thus equally easy to
analyze what happens to any real-valued $Q$.
In particular, it is straightforward to observe that
the proof of \cite{jhacm} in fact
establishes that parity of size at least 3 is
useless.  Similarly the proof of
\cite{AustrinMossel:08}, showing 
that any predicate that supports a pairwise independent measure 
is approximation resistant, also
gives uselessness (but of course we still need
to assume the unique games conjecture).

The possibly surprising, but technically not very difficult,
result that we go on to establish (in Section~\ref{sec:useful}) is that
if the condition of \cite{AustrinMossel:08} is violated
then we can find a real-valued function for which
$P$ is useful.  Thus assuming the UGC we have
a complete characterization of the property of
being useless!

\begin{theorem}
  \label{sec:main charac}
  Assuming the UGC, a predicate $P$ is (computationally) useless if and
  only if there is a pairwise independent distribution
  supported on $P^{-1}(1)$.
\end{theorem}

\paragraph{Without negated variables}

We then go on in Section~\ref{sec:noneg} to briefly discuss what
happens in the case when we do not allow negated variables, which in
some cases may be more natural.  In this situation we need to extend
the notion of a trivial algorithm in that now it might make sense to
give random but biased values to the variables.  A simple example is
when $P$ accepts the all-one string in which case setting each
variable to $1$ with probability $1$ causes us to satisfy all
constraints (regardless of the instance), but probabilities strictly between
$0$ and $1$ might also be optimal.  Taking this into account our
definitions extend.

In the setting without negated variables it turns out that the unique
games-based uselessness proof can be extended with slightly
relaxed conditions with minor
modifications.  We are still interested in a measure, $\mu$, supported
on strings accepted by $P$ but we can allow two relaxations.
The individual bits under $\mu$ need not be unbiased but each
bit should have the same bias.  Perhaps more interestingly, the bits need not be
pairwise independent and we can allow positive (but for
each pair of bits the same) correlations among the bits.

\begin{theorem}[Informal]
  When we do not allow negated variables, $P$ is useless (assuming
  UGC) if and only if the accepting strings of $P$ supports such a
  distribution.
\end{theorem}

Note that this implies that any predicate that is useless when we
allow negations is also useless when we do not allow negations
while the converse is not true.

A basic computationally useless predicate in this setting is odd parity of an
even number of variables (at least $4$ variables).  With even parity,
or with odd parity of an odd number of variables, the predicate is also
useless, but for the trivial reason that we can always satisfy all
constraints (so the guarantee that we can satisfy most applications of
$P$ gives no extra information).  Surprisingly we need the UGC to
establish the result for odd parity of an even number of variables.
As briefly discussed in Section~\ref{sec:noneg}
below it seems like new techniques
are needed to establish NP-hardness results in this situation.

\paragraph{Adaptive uselessness and pseudorandomness}
Our definition of uselessness is not the only possible choice.  A
stronger definition would be to let the algorithm choose a new
objective $Q$ based on the actual Max-$P$ instance $I$, rather than
just based on $P$.  We refer to this as \emph{adaptive} uselessness,
and discuss it in Section~\ref{sec:adaptive}.  It turns out that in
the settings discussed, adaptive uselessness is the same
as non-adaptive uselessness.

In the adaptive setting when we allow negations
clearly the task is to find an assignment
such that the $k$-bit strings appearing in the constraints
do not have the uniform distribution.  This is the
case as we can choose a $Q$ which takes large
values for the commonly appearing strings.  Thus
in this situation our results say that even given the
promise that there is an assignment such that almost
all resulting $k$-bit strings satisfy $P$, an efficient
algorithm is unable to find any assignment for which
the distribution on $k$-bit strings is not (almost) uniform.

\paragraph{Other results}

When we come to investigating useful predicates and to determining
pairs for which $P$ is useful for $Q$ it is of great value to have
extensions of the result \cite{AustrinMossel:08}.  These are along the
lines of having distributions supported on the strings accepted by $P$
where most pairs of variables are uncorrelated.  Details of this can
be found in Section~\ref{sec:generalized_am}.  Then motivated by the
pairwise independence condition we present (in Section~\ref{sec:quad})
a predicate which is the sign of quadratic function but which is still
approximation resistant.  We also take a brief look at the other end
of the spectrum, and study CSPs which are highly approximable in
Section~\ref{sec:completely approx}.

A preliminary version of this paper has appeared at
the conference for Computational Complexity \cite{ah12}.

\section{Preliminaries} \label{sec:prelim}

We have a predicate $P$: $\{ -1, 1\}^k \rightarrow \{ 0,1\}$.  The
traditional approximation problem to study is Max-$P$ in which an
instance consists of $m$ $k$-tuples of literals, each literal being
either a variable or a negated variable.  The goal is to find an
assignment to the variables so as to maximize the number of
resulting $k$-bit strings that satisfy the predicate $P$.  To be more
formal an instance is given by a set of indices $a_{j}^i \in [n]$ for
$1\leq i\leq k$ and $1\leq j\leq m$ and complementations $b_{j}^i \in
\{ -1,1\}$.  The $j$th $k$-tuple of literals contains the variables
$(x_{a_{j}^i})_{i \in [k]}$ with $x_{a_j^i}$ negated iff $b_{j}^i =-1$.  
We use the short hand
notation $P(x_{a_j}^{b_j})$ for the $j$th constraint.

We do not allow several occurrences of the same variable
in one constraint.  In other words, $a_j^i \not= a_j^{i'}$
for $i\not= i'$.  The reason for this convention is that
if the same variable appears twice we in fact have a different
predicate on a smaller number of variables.  This different
predicate is of course somewhat related to $P$ but does not share
even basic properties such as the probability that it
is satisfied by a random assignment.  Thus allowing 
repeated variables would take us into a more complicated
situation.

In this paper we
assume that all constraints have the same weight but it is not hard to
extend the results to the weighted case.  

\begin{defn}
For $Q: \{ -1, 1\}^k \rightarrow \R$
define
$$E_{Q}= \E_{x \in \{-1,1\}^k} \left[ Q(x) \right] .$$
\end{defn}

Note that for a predicate $P$ an alternative
definition of $E_{P}$ is the probability
that a uniformly random assignment satisfies $P$.
It follows that the trivial algorithm that
just picks a uniformly random assignment approximates
Max-$P$ within a factor $E_P$. 

\begin{defn}
The predicate $P$ is {\em approximation resistant}
if and only if, for every $\epsilon>0$, it is NP-hard to
approximate Max-$P$ within a factor $E_P+\epsilon$.
\end{defn}

Another way to formulate this definition is that, again for any
$\epsilon > 0$, it is NP-hard to distinguish instances for which the
optimal solution satisfies a fraction $1-\epsilon$ of the constraints
from those where the optimal solution only satisfies a fraction
$E_P+\epsilon$.  One can ask for even more and we have the following
definition.

\begin{defn}
The predicate $P$ is {\em approximation resistant
on satisfiable instances}
if and only if, for any $\epsilon>0$, it is NP-hard to
to distinguish instances of Max-$P$ for which the optimal
solution satisfies all the constraints from those 
instances where the optimal
solution only satisfies a fraction 
$E_P+\epsilon$ of the constraints.  
\end{defn}

A phenomenon that often appears is that 
if $P$ is approximation resistant then any
predicate $P'$ that accepts strictly more strings
is also approximation resistant.  Let us
introduce a concept to capture this fact.

\begin{defn}
The predicate $P$ is {\em hereditarily approximation resistant}
if and only if, for any predicate $P'$ implied by $P$ (i.e., whenever
$P(x)$ is true then so is $P'(x)$) is approximation
resistant.
\end{defn}

It turns out that 3-Lin, and indeed any parity of size at
least three, is hereditarily approximation resistant.
There are also analogous notions for satisfiable
instances but as this is not the focus of the present paper
we do not give the formal definition here.
One of the few examples of a predicate that is approximation
resistant but not hereditarily so is a predicate
studied by Guruswami et al \cite{glst}.  We discuss this predicate
in more detail in Section~\ref{sec:generalized_am} below.

%% We proceed to the key, and new, definitions of
%% this paper.

%% \begin{defn}
%% The predicate $P$ is {\em useless}
%% for the real valued function $Q$
%% $\{ -1, 1\}^k \mapsto \R$, iff
%% for any $\epsilon>0$
%% given an instance of Max-$P$ where the optimal
%% solution satisfies a fraction
%% $1-\epsilon$ of the constraints, it is NP-hard
%% to find a assignment such
%% that
%% $$
%% \frac 1m \sum_{j=1}^m Q(x_{a_j}^{b_j}) \geq E_Q+\epsilon.
%% $$
%% \end{defn}

%% We also use the term ``$P$ is useful for $Q$'' for
%% the negation of the above property.  We again
%% have the case of satisfiable instances.

%% \begin{defn}
%% The predicate $P$ is {\em useless on satisfiable instances}
%% for the real-valued function $Q$
%% $\{ -1, 1\}^k \mapsto \R$, iff
%% for any any $\epsilon>0$
%% given a completely satisfiable instance of Max-$P$,
%% it is NP-hard to find a assignment such
%% that
%% $$
%% \frac 1m \sum_{j=1}^m Q(x_{a_j}^{b_j}) \geq E_Q+\epsilon.
%% $$
%% \end{defn}

%% Since also predicates are real-valued function it is not
%% difficult to see ``$P$ is useless for 
%% $P$'' is the same as saying that $P$
%% is approximation resistant.

%% Let us give one final definition.
%% \begin{defn}
%% The predicate $P$ is {\em computationally useless} 
%% iff it is useless for any $Q$.
%% \end{defn}

%% Note that, almost by definition,  being computationally useless is a monotone
%% property.  If $P$ is computationally useless and $P$ implies $P'$ then
%% $P'$ is computationally useless and this follows by looking at
%% exactly the same instances.

Let us recall the definition of pairwise independence.

\begin{defn}
  A distribution $\mu$ over $\{-1,1\}^k$ is \emph{biased pairwise
    independent} if, for some $p \in [0,1]$, we have $\Pr_{\mu}[x_i =
    1] = p$ for every $i\in[k]$ and $\Pr_\mu[x_i = 1 \wedge x_j = 1] =
  p^2$ for every $1 \le i< j \le k$ (i.e., if all two-dimensional
  marginal distributions are equal and product distributions).

  We say that $\mu$ is \emph{pairwise independent} if it is biased
  pairwise independent with $p = 1/2$ (i.e., if the marginal
  distribution on any pair of coordinates is uniform).
\end{defn}

Finally we need a new definition of a distribution that we call
uniformly positively correlated.

\begin{defn}
  A distribution $\mu$ over $\{-1,1\}^k$ is 
    \emph{uniformly positively correlated}
    if, for some $p,\rho \in [0,1]$, with $\rho \geq p^2$,
    we have $\Pr_{\mu}[x_i =
    1] = p$ for every $i\in[k]$ and $\Pr_\mu[x_i = 1 \wedge x_j = 1] =
  \rho$ for every $1 \le i< j \le k$ (i.e., if all two-dimensional
  marginal distributions are equal and the bits are positively correlated).
\end{defn}
Note that we allow $\rho=p^2$ and thus any biased pairwise independent
distribution is uniformly positively correlated.

\section{Information-Theoretic Usefulness}\label{sec:info}

Clearly there must be some relation between $P$ and $Q$ for
our notion to be interesting and let us discuss this
briefly in the case when $Q$ is a predicate.

If $P$ and $Q$ are not strongly related then it
is possible to have instances where we can satisfy
all constraints when applying $P$ and only an $E_Q$ fraction
for $Q$.  A trivial example would be if $P$ is OR of three
variables and $Q$ is XOR.  Then
given the two constraints $(x_1,x_2,x_3)$
and $(x_1,x_2,\bar x_3)$ it is easy to satisfy
both constraints under $P$ but clearly exactly one
is always satisfied under $Q$.  Thus we conclude that OR is not useful for XOR.

As another example let $P$ be equality of two bits and $Q$
non-equality and let the constraints be all pairs $(x_i,x_j)$ for $1
\leq i < j\leq n$ (unnegated).  It is possible to satisfy all constraints under
$P$ but it is not difficult to see that the maximal fraction goes to
1/2 under $Q$ as $n$ tends to infinity.  We can note that the
situation is the same for $P$ being odd parity and $Q$ being even
parity if the size is even, while if the size of the parity is odd the
situation is completely the opposite as negating a good assignment for
$P$ gives a good assignment for $Q$.

After these examples let
us take a look in more detail at usefulness in
an information-theoretic sense. It is not
difficult to see that perfect and almost-perfect
completeness are equivalent in this situation.

\begin{defn}
A predicate $P$ is \emph{information-theoretically useless} for $Q$
if, for any $\epsilon >0$ there is an instance
such that
$$
\max_x \frac 1m \sum_{j=1}^m P(x_{a_j}^{b_j}) =1 
$$
while
$$
\max_x \frac 1m \sum_{j=1}^m Q(x_{a_j}^{b_j}) \leq E_Q+\epsilon.
$$
\end{defn}

A trivial remark is that in the information-theoretic setting
we cannot have total uselessness as $P$ is always information-theoretically
useful for itself or any predicate implied by $P$ (unless $P$ is trivial).

Let us analyze the above definition.  Let $\mu$
be a probability measure and let $\mu^p$ be the distribution
obtained by first picking a string according to $\mu$
and then flipping each coordinate with probability 
$p$.  Note that $p$ need not be small and $p=1$ is
one interesting alternative as illustrated by the parity example above.

For a given $\mu$ let $Opt(Q,\mu)$ be the maximum
over $p$ of the expected value of $Q(x)$ when $x$
is chosen according to $\mu^p$.  We have the following theorem.

\begin{theorem}
\label{thm:it-useless}
The predicate $P$ is information-theoretically useless for
$Q$ if and only if there exists a measure supported on strings
accepted by $P$ such that $Opt(Q,\mu)=E_Q$.
\end{theorem}

\begin{proof}
Let us first see that if
$Opt(Q,\mu)> E_Q$ for every $\mu$ then $P$ is indeed useful for $Q$.
Note that the space of measures on a finite set is
compact and thus we have
$Opt(Q,\mu) \geq E_Q+\delta$ for some fixed $\delta>0$ for
any measure $\mu$.

Consider any instance with
$$
\max \frac 1m \sum_{j=1}^m P(x_{a_j}^{b_j}) =1 
$$
and let us consider the strings $(x_{a_j}^{b_j})_{j=1}^m$
when $x$ is the optimal solution.  These are all accepted
by $P$ and considering with which proportion each string
appears we let this define a measure $\mu$ of strings
accepted by $P$.  By the definition of $Opt(Q, \mu)$,
there is some $p$ such that a random string from $\mu^p$
gives an expected value of at least $E_Q+\delta$ for $Q(x)$.
It follows that flipping each bit in the optimal
assignment for $P$ with probability
$p$ we get an assignment such that
$$
\E\left[\frac{1}{m} \sum_{j=1}^m Q(x_{a_j}^{b_j})\right] \geq E_Q+\delta
$$
and thus $P$ is information-theoretically useful for $Q$.

For the reverse conclusion we construct a randomized
instance.  Let $\mu$ the measure guaranteed to exist
by the assumption of the theorem.
We make sure that the all-one solution
always gives an optimal value by setting each
$a_j$ to a uniformly random set of indices from $[n]$ which are
all different, and setting $b_j$ such that the resulting
string $x_{a_j}^{b_j}$ have the distribution
given by $\mu$.

Now we claim that, for an assignment with a fraction $1-p$ variables
set to 1, the expected value (over the choice of instance) of $\frac
1m \sum_{j=1}^m Q(x_{a_j}^{b_j})$ is within an
additive $O(\frac 1n)$ of $\E[Q(x)]$ when $x$ is chosen
according to $\mu^p$.  This is more or less immediate from the
definition and the small error comes form the fact
that we require the chosen variables to be different
creating a small bias.  
Taking $m$ sufficiently large compared to $n$
the theorem now follows from standard large deviation estimates
and an application of the union bound.
\end{proof}

Let us return to our main interest of studying usefulness
in a computational context.
\section{Some Examples and Easy Theorems}
\label{sec:easyuseless}

We have an almost immediate consequence of the
definitions.

\begin{theorem}
If $P$ is useless then $P$ is hereditarily
approximation resistant.
\end{theorem}

\begin{proof}
Let $P'$ be any predicate implied by $P$.
The fact that $P$ is useless for $P'$ states
that it is hard to distinguish instances
where we can satisfy $P$ (and hence $P'$)
almost always from those where we can
only satisfy $P'$ on an $E_{P'}$ fraction 
of the constraints.  The theorem follows.
\end{proof}

Clearly we have the similar theorem for
satisfiable instances.

\begin{theorem}
If $P$ is useless on satisfiable instances then $P$ is hereditarily
approximation resistant on satisfiable instances.
\end{theorem}

The standard way to prove that a predicate $P$ is approximation
resistant is to design a Probabilistically
Checkable Proof (PCP)  where the acceptance criterion
is given by $P$ and to prove that we have almost
perfect completeness (i.e., correct proofs of correct
statements are accepted with probability $1-\epsilon$)
and soundness $E_P+\epsilon.$  Usually it is
easy to analyze the completeness and the main
difficulty is the soundness.  In this analysis of
soundness, $P$ is expanded using the discrete
Fourier transform and the expectation of each
term is analyzed separately.  

The most robust way of making this analysis is
to prove that each non-constant monomial
has expectation at most $\epsilon$.
As any real-valued function can be expanded
by the Fourier transform this argument
actually shows that the predicate in
question is computationally useless.  To show
the principle let us prove the following theorem.

\begin{theorem}
For any $k \geq 3$, parity of $k$ variables
is computationally useless.
\end{theorem}

\begin{proof}
To avoid cumbersome notation let us only give
the proof in the case $k=3$.  We assume the reader
is familiar with the
PCP defined for this case in \cite{jhacm} to
prove that Max-3-Lin is approximation resistant.
We claim that the same instances show that 
3-Lin is computationally useless.

Indeed consider an arbitrary $Q: \{-1,1\}^3 \rightarrow \R$ and
consider its Fourier-expansion
\begin{eqnarray} \label{eq: qsum}
Q(x)=\sum_{S \subseteq [3]} \hat Q_S \chi_{S}(x).
\end{eqnarray}
Now we need to consider
$
\sum_{i=1}^m Q(x_{a_j}^{b_j})
$
and we can expand each term using (\ref{eq: qsum})
and switch the the order of summation.
Remember that $\hat Q_\emptyset =E_Q$ and thus
we need to make sure that
\begin{eqnarray} \label{eq: chisum}
\sum_{i=1}^m \chi_{S}(x_{a_j}^{b_j})
\end{eqnarray}
is small for any non-empty $S$ (unless there
is a good strategy for the provers in the
underlying two-prover game).  This is
done in \cite{jhacm} for $S=\{ 1, 2, 3\}$ as
this is the only Fourier coefficient that
appears in the expansion of parity itself.

For smaller, non-empty, $S$, it is easy to see
that (\ref{eq: chisum}) equals 0.  Bits read
corresponding $A(f)$ and $B(g_i)$ are pairwise
independent and pairing terms for $f$ and
$-f$ proves that $\E[B(g_1)B(g_2)]=0$.

The result follows in the case of parity 
of 3 variables and the extension to the
general case is straightforward and left
to the reader.
\end{proof}

As stated above most approximation resistance
results turn out
to give uselessness without any or only minor modifications
of the proofs.  In particular, if one is looking
for sparse useless predicates, the predicates
by Samorodnitsky and Trevisan \cite{st} (accepting
$2^{2d}$ strings of arity $2d+d^2$) 
and of Engebretsen and Holmerin \cite{eh} (accepting
$2^{d}$ strings of arity $d(d-1)/2$) are computationally
useless.  For the former result, the proof in \cite{hw}
is easier to extend to give computational 
uselessness.

Turning to satisfiable instances, for arity 3,
the predicate 
$$
(x_1=1) \lor (x_1 \oplus x_2 \oplus x_3)
$$
studied in \cite{jhacm} is computationally useless
even on satisfiable instances.  Turning
to sparse predicates of larger arity the
predicates defined by H{\aa}stad and Khot
\cite{hkj} which accepts $2^{4k}$ inputs
and have arity $4k+k^2$, have the same property.
This paper presents two different predicates
with these parameters and although it is likely
that the result holds for both predicates we have only
verified this for ``the almost disjoint sets PCP''.
If we are willing to assume the unique games
conjecture by Khot \cite{khotunique} we can use the results
of \cite{AustrinMossel:08} to get very
strong results.

\begin{theorem}\label{thm: main hard}
Let $P$ be a predicate such that the strings accepted by
$P$ supports a pairwise independent measure.  Then,
assuming the unique games conjecture, $P$ is computationally
useless.
\end{theorem}

This follows immediately from the proof of \cite{AustrinMossel:08}
as the proof shows that the expectation of each
non-constant character is small.

As the unique games conjecture has imperfect completeness
there is no natural way to use it to prove that
certain predicates are computationally useless on
satisfiable instances.  We note, however, that
the result of O'Donnell and Wu \cite{ow:09} that 
establishes that the 3-ary predicate ``not two'' is
approximation resistant on satisfiable instances,
based on any $d$-to-1 conjecture, establishes
the same predicate to be computationally useless on
satisfiable instances.

\section{The Main Usefulness Result}\label{sec:useful}

In this section we present our main algorithm showing
that Theorem~\ref{thm: main hard} is best possible
in that any predicate that does not support a pairwise
independent measure is in fact not computationally
useless.  We have the following result which is proved in
\cite{ah11} but, as it is natural and not
very difficult,  we suspect that it is not original of that paper.

\begin{theorem}
\label{thm:pairwise quad}
Suppose that the set of inputs accepted by predicate $P$ 
does not support a pairwise independent measure.  
Then there is a real-valued quadratic polynomial $Q$ such that
$Q(x)> E_Q$ for any $x \in P^{-1}(1)$.
\end{theorem}

\begin{proof}[Proof Sketch]
The full proof appears in \cite{ah11} but let us give
a sketch of
the proof.  For each $x \in \{ -1,1\}^k$ we can define a point $x^{(2)}$
in $k+ {k\choose 2}$ real dimensions where the coordinates are
given by the coordinates of $x$ as well as any pairwise
product of coordinates $x_ix_j$.  The statement that a set $S$
supports a pairwise independent measure is equivalent with
the origin being in the convex hull of the
points
$
\{ x{^{(2)}}\ | \ x\in S\}
$.
If the origin is not in the convex hull of these points then
there is a separating hyperplane and this hyperplane defines
the quadratic function $Q$.
\end{proof}

We now have the following theorem.
\begin{theorem}\label{thm:negation useful}
Let $P$ be a predicate whose accepting inputs do not 
support a pairwise independent measure and let 
$Q$ be the quadratic function proved to exist by
Theorem~\ref{thm:pairwise quad}.  Then $P$ is useful
for $Q$.
\end{theorem}

\begin{proof}
To make the situation more symmetric let us introduce
a variable $x_0$ which always takes the value
1 and replace the linear terms $x_i$ by $x_0x_i$ and
drop any constant term in $Q$.
This makes $Q$ homogeneous of degree 2.  Note that
negating all inputs does not change the value
of $Q$ and thus any solution with $x_0=-1$ can
be transformed to a legitimate solution by
negating all variables.  As each term is unbiased
we have $E_Q=0$ and thus the goal is to find
an assignment that gives $\frac 1m \sum  Q(x_{a_j}^{b_j}) \geq \delta m$
for some absolute constant $\delta$.
Now let 
\begin{align*}
C&= \max_{x \in \{-1,1\}^k} -Q(x) &
c& = \min_{x \in P^{-1}(1)} Q(x).
\end{align*}
By assumption
we have that $c$ and $C$ are fixed constants where
$c$ is strictly larger than $0$.  Let $D$ be the
sum of the absolute values of all coefficients
of $Q$.

Let us consider our objective function
$$
F(x)=\sum_{i=1}^m Q(x_{a_j}^{b_j})
$$
which is a quadratic polynomial with
the sum of the absolute values of coefficients bounded
by $Dm$.  As we are guaranteed that we have
an assignment that satisfies at least $(1-\epsilon)m$
clauses we know that the optimal value of
$F$ is at least $(1-\epsilon) cm-\epsilon Cm \geq cm-(c+C)\epsilon m$.

Consider the standard semidefinite relaxation where
we replace each product $x_ix_j$ by an inner product
$(v_i,v_j)$ for unit length vectors $v_i$.
This semidefinite program can be solved with
arbitrary accuracy and let us for notational 
convenience assume that we have an optimal solution
which by assumption has an objective value at least
$cm-(c+C)\epsilon m$.

To round the vector valued solution back to
a Boolean valued solution we use the following
rounding guided by a positive constant $B$.

\begin{enumerate}

\item Pick a random vector $r$ by picking each
coordinate to be an independent normal variable
with mean 0 and variance 1.

\item For each $i$ if $| (v_i,r)| \leq B$ set
$p_i=\frac{B+(v_i,r)}{2B}$ and otherwise set $p_i=\frac 12$.

\item Set $x_i=1$ with probability $p_i$ independently
for each $i$ and otherwise $x_i=-1$.

\end{enumerate}
Remember that if $x_0$ gets the value $-1$ we negate
all variables.  The lemma below is the key
to the analysis.

\begin{lemma}
We have 
$$
\left|\E[x_ix_j]-\frac 1{B^2} (v_i,v_j)\right| \leq b e^{-B^2/2}.
$$
for some absolute constant $b$.
\end{lemma}

\begin{proof}
If $| (v_i,r)| \leq B$ and $| (v_j,r)| \leq B$ then
$\E[x_ix_j]= \frac 1{B^2} \E_r[(v_i,r)(v_j,r)]$.  Now
it is not difficult to see that $\E_r[(v_i,r)(v_j,r)]=(v_i,v_j)$
and thus using the fact that 
$Pr[|(v_i,r)| >B ] \leq \frac{b}{2} e^{-B^2/2}$ for a suitable
constant $b$, the lemma follows.
\end{proof}

Taking all the facts together we get that the
obtained Boolean solution has expected value
at least
$$
\frac 1{B^2} (cm-(c+C)\epsilon m) - b e^{-B^2/2}Dm.
$$
If we choose $\epsilon = \frac c{2(c+C)}$ and then
$B$ a sufficiently large constant we see that
this expected value is at least $\delta m$ for
some absolute constant $\delta$.
The theorem follows.
\end{proof}

\section{The Case of No Negation} \label{sec:noneg}

In our definition we are currently allowing negation
for free.  Traditionally this has not been the choice
in most of the CSP-literature.  Allowing negations
does make many situations more smooth but both cases
are of importance and let us here outline what
happens in the absence of negation.  We call
the resulting class Max-$P^+$.

In this case the situation is different and small
changes in $P$ may result in large difference
in performance of ``trivial'' algorithms.  In
particular, if $P$ accepts the all-zero or all-one
string then it is trivial to satisfy all constraints
by setting each variable to 0 in the first case and
each variable to 1 in the second case.

We propose to extend the set of trivial algorithms to allow the
algorithm to find a bias $r \in [-1,1]$ and then set all variables randomly with expectation $r$, independently.  The algorithm to outperform is then
the algorithm with the optimal value of $r$.  Note that this algorithm
is still oblivious to the instance as the optimal $r$ depends solely
on $P$.  We extend the definition of $E_Q$ for this setting.
\begin{defn}
For $Q$: $\{ -1, 1\}^k \mapsto \R$ and $r \in [-1,1]$, 
define
\begin{align*}
  E_{Q}(r) &= \E_{x \in \{-1,1\}^k_{(r)}} Q(x), &
  E_Q^+ &= \max_{r \in [-1,1]} E_Q(r),
\end{align*}
where $\{-1,1\}^k_{(r)}$ denotes the $r$-biased hypercube.  
\end{defn}

Using this definition we now get extensions of the definitions of
approximation resistance and uselessness of Max-$P^+$, and we say that
$P$ is \emph{positively approximation resistant} or \emph{positively
  useless}.

\subsection{Positive usefulness in the information
theoretic setting}\label{sec:info+}

The results of Section~\ref{sec:info} are not difficult
to extend and we only give an outline.  The main new component
to address is the fact that 0 and 1 are not symmetric any longer.

As before let $\mu$
be a probability measure and let $\mu^{p,q}$ be the distribution
obtained by first picking a string according to $\mu$
and then flipping each coordinate that is one to a zero with
with probability $p$ and each coordinate that is zero to one with probability
$q$ (of course all independently).
For a given $\mu$ let $Opt^+(Q,\mu)$ be the maximum
over $p$ and $q$ of the expected value of $Q(x)$ when $x$
is chosen according to $\mu^{p,q}$.  We have the following theorem.

\begin{theorem}
\label{thm:it-useless+}
The predicate $P$ is positively information-theoretically useless for
$Q$ if and only if there exists a measure supported on strings
accepted by $P$ such that $Opt^+(Q,\mu)=E_Q^+$.
\end{theorem}

\begin{proof}
The proof follows the proof of Theorem~\ref{thm:it-useless},
and we leave the easy modifications to the reader.
\end{proof}
Let us return to the more interesting case of studying
positive uselessness in the computational setting.

\subsection{Positive usefulness in the computational setting}

Also in this situation we can extend the result from
the situation allowing negations by using very similar
techniques.
We first extend the hardness result Theorem~\ref{thm: main hard} based on
pairwise independence to this setting and we can now even allow a
uniformly positively correlated distribution.

\begin{theorem}\label{thm:no neg hard}
Let $P$ be a predicate such that the strings accepted by $P$ supports
a uniformly positively correlated distribution.  Then, assuming the
unique games conjecture, $P$ is positively useless.
\end{theorem}

A similar theorem was noted in \cite{a-submodular10}, but that theorem
only applied for pairwise independent distributions.  The relaxed
condition that the distribution only needs to be positively correlated
is crucial to us as it allows us to get a tight characterization.
As the proof of Theorem~\ref{thm:no neg hard} has much in common
with the proof of Theorem~\ref{thm:resist_general} stated below
we give the proofs of both theorems in Section~\ref{sec:ugproofs}.

Let us turn to establishing the converse of Theorem~\ref{thm:no neg hard}.
We start by extending Theorem~\ref{thm:pairwise quad}.

\begin{theorem}
\label{thm:positive quad}
Suppose that the set of inputs accepted by predicate $P$ 
does not support a uniformly positively correlated measure.  
Then there is a real-valued quadratic polynomial $Q$ such that
$Q(x)> E_Q^+$ for any $x \in P^{-1}(1)$.  Furthermore, $Q$ can
be chosen such that the 
optimal bias $r$ giving the value $E_Q^+$ satisfies
$|r| < 1$.
\end{theorem}

\begin{proof}
As in the proof of Theorem~\ref{thm:pairwise quad} for each $x \in \{
-1,1\}^k$ we can define a point $x^{(2)}$ in $k+ {k\choose 2}$ real
dimensions where the coordinates are given by the coordinates of $x$
as well as any pairwise product of coordinates $x_ix_j$.  We consider
two convex bodies, $K_1$ and $K_2$ where $K_1$ is the same body we saw in
the proof of Theorem~\ref{thm:pairwise quad} -- the convex hull of
$x^{(2)}$ for all $x$ accepted by $P$.

For each $b \in [ -1,1]$ we have a point $y^b$ with
the first $k$ coordinates equal to $b$ and the rest
of the coordinates equal to $b^2$.  We let $K_2$ be the
convex hull of all these points.

The hypothesis of the theorem
is now equivalent to the fact that $K_1$ and $K_2$ are
disjoint.  Any hyperplane separating these two
convex sets would be sufficient for the first part of
the theorem but to make sure that the optimal $r$
satisfies $|r|<1$ we need to consider how to find this
hyperplane more carefully.

Suppose $p_2$ is a point in $K_1$ such that $d(p_2,K_1)$, i.e., the distance
from $p_2$ to $K_1$, is minimal.  Furthermore let $p_1$ be
the point in $K_1$ minimizing $d(p_1,p_2)$.  One choice for the
separating hyperplane is the hyperplane which is orthogonal to the
line through $p_1$ and $p_2$ and which intersects this line
at the midpoint between $p_1$ and $p_2$.
We get a corresponding quadratic form, $Q$, and it is not difficult
to see that the maximum of $Q$ over $K_2$ is taken at $p_2$
(and possibly at some other points).
Thus if we can make sure that $p_2$ does not equal
$(1^k, 1^{k \choose 2})$ or $(-1^k, 1^{k \choose 2})$ we are
done.  

We make sure that this is the case by first
applying a linear transformation to the space.
Note that applying a linear transformation does not
change the property that $K_1$ and $K_2$ are non-intersecting
convex bodies but it does change the identity of the
points $p_1$ and $p_2$.

As $P$ does not support a uniformly positively correlated measure
it does not accept either of the points $1^k$ or $-1^k$
as a measure concentrated on such a point is
uniformly positively correlated.
This implies that $K_1$ is contained in the
strip
$$
\left|\sum_{i=1}^k y_i \right| \leq k-2.
$$
We also have that $K_2$ is contained in the strip
$$
\left|\sum_{i=1}^k y_i \right| \leq k,
$$
and that it contains points with the given sum taking
any value in the above interval.
Furthermore the points we want avoid satisfy
$|\sum_{i=1}^k y_i|=k$.  Now apply a linear transformation
that stretches space by a large factor in the
direction of the vector $(1^k,0^{k \choose 2})$ while preserving
the space in any direction orthogonal to this vector.
It is easy to see that for a large enough stretch factor,
none of the points $(1^k, 1^{k \choose 2})$ or $(-1^k, 1^{k \choose 2})$ 
can be the point in $K_2$ that is closest to $K_1$.
The theorem follows.
\end{proof}

Given Theorem~\ref{thm:negation useful} the next
theorem should be no surprise.

\begin{theorem}\label{thm:negation useful+}
Let $P$ be a predicate whose set of accepting inputs
does not support a uniformly positively correlated measure and
$Q$ be the quadratic function proved to exist by
Theorem~\ref{thm:positive quad}.  Then $P$ is positively useful
for $Q$.
\end{theorem}

\begin{proof}
The proof is small modification of the proof of
Theorem~\ref{thm:negation useful} and let us only
outline these modifications.

Let $r$ be the optimal bias of the inputs to get
the best expectation of $Q$ and let us consider
the expected value of $\frac 1m \sum Q(x_{a_j}^{b_j})$ given that we set
$x_i$ to one with probability $(1+r+y_i)/2$.  This probability can
be written a quadratic form in $y_i$ and we want to
optimize this quadratic form under the conditions that $|r+y_i|=1$
for any $i$.
Note that the constant term is $E_Q^+$ and if we introduce
a new variable $y_0$ that always takes the value 1 we can
write the resulting expectation as
\begin{eqnarray}\label{acc prob}
E_Q^+ + \sum_{i\not=j} c_{ij} y_i y_j,
\end{eqnarray}
for some real coefficients $c_{ij}$.
As before we relax this to a semi-definite
program by replacing the products $y_iy_j$ in (\ref{acc prob}) 
by inner products $(v_i,v_j)$ and relaxing the constraints to
$$\| rv_0 + v_i\| \leq 1 ,$$
for any $i\geq 1$ and $\|v_0\|=1$.  
Solving this semi-definite program we
are now in essentially the same situation as in 
the proof of Theorem~\ref{thm:negation useful}.
The fact that $|r|<1$ ensures
that a sufficiently large scaling of the inner products
results in probabilities in the interval $[0,1]$.
We omit the details.
\end{proof}

Theorem~\ref{thm:no neg hard} proves that having odd parity
on four variables is positively useless but assumes the UGC.
It would seem natural that this theorem should be possible
to establish based solely on $NP\not= P$, but we have been
unable to do so.  

Let us briefly
outline the problems encountered.  The natural attempt is to try
a long-code based proof for label cover instance
similar to the proof \cite{jhacm}.  A major problem seems to
be that all known such proofs read two bits from the same
long-code.  Considering functions $Q$ that benefit from two
such bits being equal gives us trouble through incorrect
proofs where each individual long code is constant. 
For instance we currently do not know how to show that
odd parity is not useful for the ``exactly three'' function
based only on NP$\not=$P.

\section{Adaptive Uselessness and Pseudorandomness}
\label{sec:adaptive}

We now discuss the adaptive setting, when we allow the algorithm to
choose the new objective function $Q$ based on the Max-$P$ instance.
Formally, we make the following definition.

\begin{defn}
The predicate $P$ is {\em adaptively useful}, if
and only if there is an $\epsilon>0$ such that there is a polynomial time algorithm which given a Max-$P$ instance with value $1-\epsilon$ finds an objective function $Q: \{-1,1\}^k \rightarrow [0,1]$ and an assignment $x$ such that
$$
\frac 1m \sum_{j=1}^m Q(x_{a_j}^{b_j}) \geq \E_{x \in \{-1,1\}^k}[Q(x)] +\epsilon.
$$
\end{defn}

Note that we need to require $Q$ to be bounded since otherwise the
algorithm can win by simply scaling $Q$ by a huge constant.
Alternatively, it is easy to see that in the 
presence of negations adaptive usefulness is
equivalent with requiring that the algorithm finds an assignment $x$
such that the distribution of the $k$-tuples $\{x_{a_j}^{b_j}\}_{j \in
  [m]}$ is $\epsilon$-far in statistical distance from uniform for some
$\epsilon > 0$ (not the same $\epsilon$ as above).  In fact, since $k$
is constant it is even equivalent with requiring that the min-entropy
is bounded away from $k$, in particular that there is some $\alpha \in
\{-1,1\}^k$ and $\epsilon > 0$ such that at least a $2^{-k}+\epsilon$
fraction of the $x_{a_j}^{b_j}$'s attain the string $\alpha$.

Adaptive uselessness trivially implies non-adaptive uselessness.  In
the other direction, with the interpretation of avoiding uniform
$k$-tuples, it is easy to see that the proof of the hardness result
based on pairwise independence from the non-adaptive setting works
also for adaptive uselessness.  

This result can, by a slightly more careful argument, be extended also
to the case without negations.  The characterization is then that the
algorithm is attempting to produce a distribution on $k$-tuples that
is far from being uniformly positively correlated.
In this setting, it does not seem meaningful to think of adaptive uselessness as
a pseudorandomness property.  

\section{Some New Approximation Resistance Results}

In this section we provide two new results on approximation
resistance.  First, we describe how the pairwise independence
condition of \cite{AustrinMossel:08} can be relaxed somewhat to give
approximation resistance for a wider range of predicates.  Second,
motivated by Theorem~\ref{thm:pairwise quad}, we show that there exist
predicates $P$ which are of the form $\sgn(Q)$ for a quadratic form
$Q$.

\subsection{Relaxed Pairwise Independence Conditions}
\label{sec:generalized_am}

Let us first recall one of the few known examples of
a predicate that is approximation resistant but
not hereditarily approximation resistant.

\begin{example}
  \label{example:glst}
  Consider the predicate $GLST: \{-1,1\}^4 \rightarrow \{0,1\}$ defined by
  $$
  GLST(x_1, x_2, x_3, x_4) = \left\{\begin{array}{ll}
      x_2 \ne x_3 & \text{if $x_1 = -1$}\\
      x_2 \ne x_4 & \text{if $x_1 = 1$}
    \end{array}\right..
  $$ 
  This predicate was shown to be approximation resistant by Guruswami
  et al.\ \cite{glst}, but there is no pairwise independent
  distribution supported on its accepting assignments -- indeed it is
  not difficult to check that $x_2x_3+x_2x_4+x_3x_4 <0$ for all
  accepting inputs.  This predicate also implies $NAE(x_2,x_3,x_4)$,
  the not-all-equal predicate and this is known to be non-trivially
  approximable \cite{z3}.

  When the predicate $GLST$ is proved to be approximation resistant in
  \cite{glst} the crucial fact is that not all terms appear in the
  Fourier expansion of $P$.   We have
  $$GLST(x_1, x_2, x_3, x_4) = \frac{1}{2} - \frac{x_2x_3}{4} -
  \frac{x_2x_4}{4} + \frac{x_1x_2x_3}{4} - \frac{x_1x_2x_4}{4}.$$ The
  key is that no term in the expansion contains both of the variables
  $x_3$ and $x_4$, corresponding to two questions in the PCP that are
  very correlated and hence giving terms that are hard to control.

%%   Consider the uniform distribution over
%%   $\{\,(x_1,x_2,x_3,x_4)\,:\,x_1x_2x_3 = -1 \wedge x_3 = -x_4\,\}$.
%%   It is not hard to see that this distribution is supported on
%%   $P^{-1}(1)$.
\end{example}

In other words, when proving approximation resistance it suffices to
only analyze those terms appearing in the Fourier expansion of a
predicate $P$.  In the context of the pairwise independent condition
(which only gives UG-hardness, not NP-hardness), this means that it
suffices to find a distribution which is pairwise independent on those
pairs of variables that appear together in some term.

However, these are not the only situations where we can deduce that
$P$ is approximation resistant.

\begin{example}
  \label{example:glst+}
  Consider the predicate $$P(x_1, x_2, x_3, x_4) = GLST(x_1, x_2, x_3,
  x_4) \vee (x_1 = x_2 = x_3 = x_4 = 1),$$ which is the $GLST$
  predicate with the all-ones
  string as an additional accepting assignment.  One can check that
  there is no pairwise independent distribution supported on
  $P^{-1}(1)$, and since $P$ has an odd number of accepting
  assignments, all its Fourier coefficients are non-zero.  However,
  Max-$P$ is known to be approximation resistant
  \cite{hast}.
\end{example}

The result of \cite{hast} proving that this predicate is
resistant is somewhat more general.  In particular, it says the
following.

\begin{theorem}[\cite{hast}, Theorem 6.5]
  \label{thm:hast_hardness}
  Let $P: \{-1,1\}^4 \rightarrow \{0,1\}$ be a predicate on $4$ bits.
  Suppose $\hat{P}(\{3,4\}) \ge 0$ and that $P$ accepts all strings
  $x_1x_2x_3x_4$ with $\prod_{i=1}^3 x_i = -1$ and $x_3 = -x_4$.
  Then $P$ is approximation resistant.
\end{theorem}

The statement of \cite{hast}, Theorem 6.5 is slightly different.  The
above statement is obtained by flipping the sign of $x_3$ in the
statement of \cite{hast}.  We now give a natural generalization of
Theorem~\ref{thm:hast_hardness} to a much larger class of predicates
(but giving UG-hardness, whereas \cite{hast} gives NP-hardness).  We
first define the specific kind of distributions whose existence give
our hardness result.

\begin{defn}
  A distribution $\mu$ over $\{-1,1\}^k$ \emph{covers} $S
  \subseteq [k]$ if there is an $i \in S$ such that $\E_\mu[x_i] = 0$
  and $\E_\mu[x_ix_j] = 0$ for every $j \in S \setminus \{i\}$.
\end{defn}

\begin{defn}
  Fix a function $Q: \{-1,1\}^k \rightarrow \R$ and a pair of
  coordinates $\{i,j\} \subseteq [k]$.  We say that a distribution
  $\mu$ over $\{-1,1\}^k$ is $\{i,j\}$-negative with respect to $Q$ if
  $\E_\mu[x_i] = \E_\mu[x_j] = 0$ and $\Cov_\mu[x_i,x_j] \hat{Q}(\{i,j\}) \le 0$.
\end{defn}

Our most general condition for approximation resistance (generalizing
both Theorem~\ref{thm:hast_hardness} and \cite{AustrinMossel:08}) is
as follows.

\begin{theorem}
  \label{thm:resist_general}
  Let $P: \{-1,1\}^k \rightarrow \{-1,1\}$ be a predicate
  and let $Q: \{-1,1\}^k \rightarrow \R$ be a real valued
  function.  Suppose
  there is a probability distribution $\mu$ supported on $P$ with the
  following properties:
  \begin{itemize}
  \item For each pair $\{i,j\} \subseteq [k]$, it holds that $\mu$ is
    $\{i,j\}$-negative with respect to $Q$
  \item For each $S \ne \emptyset, |S| \ne 2$ such that $\hat{Q}(S)
    \ne 0$, it holds that $\mu$ covers $S$
  \end{itemize}
  Then $P$ is not useful for $Q$, assuming the Unique
  Games Conjecture.  In particular, if the conditions are true
  for $Q=P$ then $P$ is approximation resistant.
\end{theorem}

We give the proof of the theorem in Section~\ref{sec:ugproofs}.
Let us now illustrate it by applying it to the earlier example.

\begin{example}[Example~\ref{example:glst+} continued]
  Consider the distribution $\mu$ used to prove approximation
  resistance for $GLST$, i.e., the uniform distribution over the four
  strings $x_1x_2x_3x_4$ satisfying $x_1x_2x_3 = -1$ and $x_3 = -x_4$
  (note that the condition of Theorem~\ref{thm:hast_hardness} is
  precisely that $P$ should accept these inputs).  First, it satisfies
  $$\hat{P}(\{3,4\})\E_\mu[x_3x_4] = \frac{1}{16} \cdot (-1) < 0,$$
  and all other pairwise correlations are $0$, so $\mu$ satisfies the
  $\{i,j\}$-negativity condition of Theorem~\ref{thm:resist_general}.
  Further, for $|S| > 2$ it holds that either $x_1$ or $x_2$ is in
  $S$.  Since $\E_\mu[x_1] = 0$ and $\E_\mu[x_1x_j] = 0$ for all $j
  \ne 1$ (and similarly for $x_2$), this shows that any $|S|>2$ is
  covered by $\mu$.  Finally since all $\E_\mu[x_i] = 0$, all four
  singleton $S$ are also covered by $\mu$.  Hence
  Theorem~\ref{thm:resist_general} implies that Max-$P$ is
  resistant (under the UGC).
\end{example}

\begin{example}
  Consider the predicate $P(x) = x_1 \oplus ((x_2 \oplus x_3) \vee
  x_4)$.
  This predicate is known to be approximation resistant \cite{hast}.
  Let us see how to derive this conclusion using
  Theorem~\ref{thm:resist_general} (albeit only under the UGC).  The
  Fourier expansion of $P$ is
  $$ 
  P(x) = \frac{1}{2} + \frac{x_1}{4} - \frac{x_1x_4}{4} - \frac{x_1x_2x_3}{4} - \frac{x_1x_2x_3x_4}{4},
  $$ 
  and the distribution we use is uniform over:
  $$\{\,x \in \{-1,1\}^4\,|\,x_1 x_2 x_3 = -1, x_4 = 1\}.$$ Each of
  $x_1, x_2, x_3$ are unbiased, and $x_4$ is completely biased but as
  it does not appear as a singleton in the expansion of $P$ this is
  not an issue.  Further, all pairwise correlations are $0$, and it is
  easy to check that this is sufficient for
  Theorem~\ref{thm:resist_general} to apply.
\end{example}

We have only used Theorem~\ref{thm:resist_general} to get
approximation resistance in a few examples but it can also
be used to give examples of $P$ not being useful for $Q$ 
in various situations but
we leave the creation of interesting such examples to 
the reader.

We are not aware of any approximation resistant predicates that do not
satisfy the conditions given in Theorem~\ref{thm:resist_general}.
On the other hand we see no reason to believe
that it is tight.

\subsection{Resistant Signs of Quadratic Forms}\label{sec:quad}

Let us consider a slightly different example. 
Suppose that in the definition of uselessness
we only considered predicates $Q$ rather
than arbitrary real-valued functions.  Would we get
the same set of useless predicates?

The answer to this question is not obvious.  Any real-valued function
$Q$ can be written in the
form
$$
Q(x)= q_0 + \sum_P c_P P(x)
$$ where the sum is over different predicates and each coefficient
$c_P$ is non-negative.  This implies that if $P$ is useful for a
real-valued function $Q$ then there is a collection of predicates such
that on any instance we can do better than random on one of these
predicates.  This does not imply that there is a single predicate for
which $P$ is useful but it excludes the standard proofs where the
instance used to prove that $P$ is useless for $Q$ is independent of
the identity of $Q$.

If one would single out a candidate predicate for which
$P$ is useful the first candidate that comes to mind
given the discussions of Section~\ref{sec:useful} is, possibly,
$$
P'(x)=\sgn(Q(x))
$$
where $Q$ is the quadratic form guaranteed by
Theorem~\ref{thm:pairwise quad}.  Note
that it may or may not be the case that $P=P'$.  This choice does not
always work.

\begin{theorem}
\label{thm:quadsignresist}
There is a predicate, $P$, of the form $\sgn(Q(x))$
where $Q$ is a quadratic function without a constant
term that is approximation resistant (assuming the UGC).
\end{theorem}

\begin{proof}
Let $L_1$ and $L_2$ be two linear
forms with integer coefficients which only
assume odd values and only depends on
variables $x_i$ for $i \geq 3$.  Define
\begin{eqnarray}\label{eq qdef}
Q(x)= 10 (L_1(x)+x_1)(L_2(x)+x_2)+x_1 L_2(x)+2 x_2 L_1(x),
\end{eqnarray}
and let $P(x)=\sgn(Q(x))$.  We establish the
following properties of $P$.

\begin{enumerate}

\item For all $\alpha$ such that $\{ 1, 2\} \subseteq \alpha$
  we have $\hat P_\alpha=0$.

\item There is a probability distribution $\mu$ supported
  on strings accepted by $P$ such that  $\E_\mu[x_i] = 0$
  for all $i$ and $\E_\mu [x_ix_j]=0$ for all
  $i< j$ with $(i,j)\not= (1,2)$.

\end{enumerate}

These two conditions clearly makes it possible to 
apply Theorem~\ref{thm:resist_general}.
Loosely speaking the
second condition makes it possible to construct
at PCP such that we can control sums over
all nontrivial characters except those that contain
both 1 and 2.  The first conditions implies that
these troublesome terms do not appear in the
expansion of $P$ and hence we can complete
the analysis.  

We claim that property 1 is equivalent to the
statement that every setting of the variables $x_i$
for $i \geq 3$ results in a function on $x_1$ and $x_2$ 
that has the Fourier coefficient of size 2 equal to
0.  In other words it should be a constant, one
of the variables $x_1$ or $x_2$ or the negation of
such a variable. Let us check that this is the
case for $Q$ defined by (\ref{eq qdef}).

Fix any value of $x_i$, $i \geq 3$ and we have
the following cases.

\begin{enumerate}

\item $|L_1| \geq 3$ and $|L_1| \geq 3$.

\item $|L_1| =1$ and $|L_2| \not= 1$.

\item $|L_2| =1$.

\end{enumerate}

In first case clearly the first term determines the
sign of $Q$ and $P=\sgn(L_1(x)L_2(x))$ and
in particular the sub-function is independent of $x_1$
and $x_2$.

The second case is almost equally straightforward.
When $x_1= L_1(x)$ then the first term dominates
and the answer is $\sgn(x_1L_2(x))$.
When  $x_1= -L_1(x)$ the first term is 0 and
as $|L_2(x) x_1| \geq 3$ while 
$|L_1(x) x_2|=1$ the answer also in this case
is $\sgn(x_1L_2(x))$.

Finally let us consider the third case.  Then if
$x_2=L_2(x)$ our function $Q$ reduces to
$$
20 (L_1(x)+x_1) x_2 + x_2(2L_1(x)+x_1)
$$
and any nonzero term of this sum has sign
$\sgn(x_2L_1(x))$.  Finally if $x_2=-L_2(x)$ we
get
$$
x_2(2L_1(x)-x_1)
$$
and again the sign is that of
$\sgn(x_2L_1(x))$. 
We conclude that in each case we have one of
the desired functions and property 1 follows.

We establish property 2 in the case when
each $L_i$ is the sum of 5 variables not occurring
in the other linear form.  Thus for example
we might take
$$
L_1(x)=x_3+x_4+x_5+x_6+x_7
$$
and
$$
L_2(x)=x_8+x_9+x_{10}+x_{11}+x_{12}.
$$
We describe the distribution $\mu$ in a rather
indirect way to later be able to analyze it.
Let $c=\frac {7-\sqrt{41}}{8}\approx .0746$.

\begin{enumerate}

\item Fix $|L_1(x)|$ to 1,3 or 5 with probabilities
$\frac 12+2c$, $\frac 12-3c$, and $c$,
respectively.

\item Fix $|L_2(x)|$ to 1 or 3 each with a probability
$\frac {1}{2}$.

\item Pick a random $b\in \{-1, 1\}$ taking each value
with probability $\frac {1}{2}$.

\item  Suppose $|L_1(x)| \geq 3$ and $|L_2(x)| \geq 3$.
Set $\sgn(L_1(x))=\sgn(L_2(x))=b$
and $x_1=x_2=-b$.

\item Suppose $|L_2|\not=1$ and $|L_1|=1$.
Set $\sgn(L_1(x))=-\sgn(L_2(x))=b$
and $x_1=x_2=-b$.

\item Suppose $|L_2| =1$. Set $\sgn(L_1(x))=x_1=x_2=b$
and $\sgn(L_2(x)))=-b$ with probability $\frac {1+12c}{2}$
and $\sgn(L_2(x))=b$ with probability $\frac {1-12c}{2}$.

\item Choose $x_i$ for $i \geq 3$ uniformly at random
given the values $L_1(x)$ and $L_2(x)$.

\end{enumerate}

Now first note that by the analysis in establishing
property 1 we always pick an assignment such that
$Q(x)>0$.  This follows as in the three
cases the output of the function is
$\sgn(L_1(x)L_2(x))$,
$\sgn(x_1L_2(x))$, and
$\sgn(L_1(x)x_2)$, respectively and they are
all chosen to be $b^2$.

As $b$ is a random bit, it is easy to see that
$\E[x_i]=0$ for any $i$ and we need to analyze
$\E[x_ix_j]$ for $(i,j)\not= (1,2)$.   In our
distribution we always have $x_1=x_2$ and the
variables in $L_1$ and $L_2$ are treated symmetrically
and hence it is sufficient to establish
the following five facts.
\begin{enumerate}
\item $\E[L_1^2(x)]=5$.
\item $\E[L_2^2(x)]=5$.
\item $\E[x_1 L_1(x)]=0$.
\item $\E[x_1 L_2(x)]=0$.
\item $\E[L_1(x) L_2(x)]=0$.
\end{enumerate}
The first expected value equals
$$
\left(\frac 12+2c\right)\cdot 1+\left(\frac 12-3c\right)\cdot 9+25 \cdot c=5
$$
while the second equals
$$
\frac {1\cdot 1 + 1 \cdot 9 }{2}=5.
$$
For the third expected value note that $x_1L_1(x)= -| L_1(x)|$
when $L_2(x)=3$ while it equals
$x_1L_1(x)= | L_1(x)|$ when $L_2(x)=1$.  The two cases
happens each with probability $1/2$ and as $| L_1(x)|$
is independent of $|L_2(x)|$ the equality follows.

To analyze the fourth value first observe that
conditioned on $|L_2(x)|=1$ we have $\E[x_1 L_2(x)]=-12c$.
On the other hand when $|L_2(x)|=3$ we have 
$$\E[x_1 L_2(x)]= 3 (\frac 12+2c)-3(\frac 12-3c)-3c=12c,
$$
giving the result in this case.  Finally, conditioned on
$|L_2(x)|=1$ we have 
$$\E[L_1(x) L_2(x)]=-12c\left( (\frac 12+2c)+3(\frac 12-3c)+5c\right)=-(24c-24c^2)$$
and
conditioned on
$|L_2(x)|=3$ we have 
$$\E[L_1(x) L_2(x)]=-3(\frac 12+2c)+9(\frac 12-3c)+15c=3-18c$$
giving a total expected
value of 
$$
3+24c^2-42c
$$
and $c$ was chosen carefully to make this
quantity 0.
\end{proof}

\section{One Result at the Other End of the Spectrum}
\label{sec:completely approx}

We have focused on computationally useless predicates that
do not enable us to do essentially anything.  Knowing
that there is an assignment that satisfies almost
all the conditions does not enable us to do better 
for any function.

At the other end of the spectrum we could hope for predicates where
even more moderate promises can be sufficient to find useful
assignments efficiently.

One possibility is to ask for a predicate that is useful
for all functions $Q$.  This is too much to ask for, as discussed
in Section~\ref{sec:info}, if $P$ and $Q$ are sufficiently
unrelated it might be the case that there are instances
where we can satisfy $P$ on all constraints while the best
assignment when we consider condition $Q$ only 
satisfies essentially a fraction $E_Q$.  One possible
definition is to say that $P$ should be useful for any
$Q$ which is not excluded by this information theoretic
argument.  This is a potential avenue of research which we
have not explored and hence we have no strong feeling about
what to expect.  One complication here is of course that
the characterization of Theorem~\ref{thm:it-useless} is not very
explicit and hence might be difficult to work with.

The $Q$ we must always consider is the traditional
question of approximability namely $Q=P$ but let us weaken
the promise from the optimum being almost one to
being just slightly above the random threshold.

\begin{defn}
A predicate $P$ is {\em fully approximable}
if for any $\epsilon >0$ there is a $\delta>0$ such
that if the optimal value of a Max-$P$ instance
is $E_P+\epsilon$ then one can efficiently
find an assignment that satisfies an $(E_P+\delta)$-fraction
of the constraints.
\end{defn}

First note that the most famous example of a fully
approximable predicate is Max-Cut and in fact
any predicate of arity two is fully approximable.
This definition has been explored previously in \cite{jhbcc}
but given that this is not a standard venue for
results on Max-CSPs let us restate the theorem which
that any fully approximable predicate is in fact
a real valued sum of predicates of arity two.

\begin{theorem}
\cite{jhbcc} A predicate $P$ is fully approximable if and only if
the Fourier expansion of $P$ contains no
term of degree at least 3.
\end{theorem}

We refer to \cite{jhbcc} for the not too difficult proof.
It is an amusing exercise to find the complete
list of such predicates.  A predicate on three variables
is fully approximable iff it accepts equally many
even and odd strings.  Up to negations and permutations of variables, the only predicate that
depends genuinely on four variables with this
property is
$$
P(x)=\frac{2+x_1x_3+x_1x_4+x_2x_3-x_2x_4}{4}.
$$

\section{Proofs of UG-Hardness}\label{sec:ugproofs}

In this section we give the proofs of the extensions
Theorems~\ref{thm:no neg hard} and \ref{thm:resist_general} of
\cite{AustrinMossel:08}.  It is well-known that the key part in
deriving UG-hardness for a CSP is to design \emph{dictatorship tests}
with appropriate properties --- see e.g. \cite{raghavendra} for details.

\subsection{Background: Polynomials, Quasirandomness and Invariance}

To set up the dictatorship test we need to mention some background
material.

For $b \in [-1,1]$, we use $\{-1,1\}^n_{(b)}$ to denote the
$n$-dimensional Boolean hypercube with the $b$-biased product
distribution, i.e., if $x$ is a sample from $\{-1,1\}^n_{(b)}$ then
the expectation of $i$'th coordinate is $\E[x_i] = b$ (equivalently, $x_i = 1$ with probability $(1+b)/2$),
independently for each $i \in [n])$.  Whenever we have a function $f:
\{-1,1\}^n_{(b)} \rightarrow \R$ we think of it as a random variable
and hence expressions like $\E[f]$, $\Var[f]$, etc, are interpreted as
being with respect to the $b$-biased distribution.  We equip $L^2(\{-1,1\}^n_{(b)})$ with the inner product
$\scalprod{f,g} = \E[f \cdot g]$ for $f, g: \{-1,1\}^n_{(b)} \rightarrow \R$.

For $S \subseteq [n]$ define $\chi_S: \{-1,1\}^n_{(b)} \rightarrow \R$ by 
$$\chi_S(x) = \prod_{i \in S} \chi(x_i),$$
where $\chi: \{-1,1\}_{(b)} \rightarrow \R$ is defined by 
$$\chi(x_i) = \frac{x_i - \E[x_i]}{\sqrt{\Var[x_i]}} =  \left\{
\begin{array}{ll} 
  -\sqrt{\frac{1+b}{1-b}} & \text{if $x_i = -1$}\\
  \sqrt{\frac{1-b}{1+b}} & \text{if $x_i = 1$}
\end{array}\right..
$$
The functions $\{\chi_S\}_{S \subseteq [n]}$ form an orthonormal basis with respect to the inner product $\scalprod{\cdot, \cdot}$
on  $L^2(\{-1,1\}^n_{(b)})$ and thus any function $f: \{-1,1\}^n_{(b)} \rightarrow \R$ can
be written as
$$
f(x) = \sum_{S \subseteq [n]} \hat{f}(S; b) \chi_S(x),
$$
where $\hat{f}(S; b)$ are the Fourier coefficients of $f$ (with respect
to the $b$-biased distribution).

With this view in mind it is convenient to think of functions $f$
in $L^2(\{-1,1\}^n_{(b)})$ as multilinear polynomials $F: \R^n
\rightarrow \R$ in the random variables $X_i = \chi(x_i)$, viz.,
$$F(X) = \sum_{S \subseteq [n]} \hat{f}(S; b) \prod_{i \in S} X_i.$$

We say that such a polynomial is \emph{$(d, \tau)$-quasirandom} if for
every $i \in [n]$ it holds that
$$
\sum_{\substack{i \in S \subseteq [n] \\ |S| \le d}} \hat{f}(S;b)^2 \le \tau.
$$ A function $f: \{-1,1\}^n_{(b)} \rightarrow \R$ is said to a be a
\emph{dictator} if $f(x) = x_i$ for some $i \in [n]$, i.e., $G$ simply
returns the $i$'th coordinate.  The polynomial $F$ corresponding to a
dictator is $F(X) = b + \sqrt{1-b^2}X_i$.
Note that a dictator is in some sense the extreme opposite of a
$(d,\tau)$-quasirandom function as a dictator is not even
$(1,\tau)$-quasirandom for any $\tau < 1$.

We are interested in distributions $\mu$ over $\{-1,1\}^k$.
In a typical situation we pick $n$ independent samples of $\mu$, resulting in
$k$ strings $\vec x_1, \ldots, \vec x_k$ of length $n$, and 
to each such string we apply 
some function $f: \{-1,1\}^n \rightarrow \{-1,1\}$.  With
this in mind, define the following $k \times n$ matrix $X$ of random
variables.  The $j$'th column which we denote by $X^j$ has the
distribution obtained by picking a sample $x \in \{-1,1\}^k$ from
$\mu$ and letting $X^j_i = \chi(x_i)$, independently for each $j \in [n]$.
Then, the distribution of $(f(\vec x_1), \ldots, f(\vec x_k))$ is the
same as the distribution of $(F(X_1), \ldots, F(X_k))$, where $X_i$
denotes the $i$'th row of $X$.

Now, we are ready to state the version of the invariance principle
\cite{MoOdOl:10,Mossel:10} that we need.
\begin{theorem}
  \label{thm:invariance}
  For any $\alpha > 0, \epsilon > 0, b \in [-1,1], k \in \mathbb{N}$
  there are $d, \tau > 0$ such that the following holds.  Let $\mu$ be any distribution over $\{-1,1\}^k$ satisfying:
  \begin{enumerate}
  \item $\E_{x \sim \mu}[x_i] = b$ for every $i \in [k]$ (i.e., all biases are identical).
  \item $\mu(x) \ge \alpha$ for every $x \in \{-1,1\}^k$ (i.e., $\mu$ has full support).
  \end{enumerate}
  Let $X$ be the $k \times n$ matrix defined above, and let $Y$ be a
  $k \times n$ matrix of standard jointly Gaussian variables with the
  same covariances as $X$.  Then, for any $(d,\tau)$-quasirandom
  multilinear polynomial $F: \R^n \rightarrow \R$, it holds that
  $$
  \left|\E\left[\prod_{i = 1}^k F(X)\right] - \E\left[\prod_{i=1}^k F(Y)\right]\right| \le \epsilon.
  $$
\end{theorem}

\subsection{The Dictatorship Test}

The dictatorship tests we use to prove
Theorems~\ref{thm:no neg hard} and \ref{thm:resist_general} are both
instantiations of the test used in \cite{AustrinMossel:08}, with
slightly different analyses, so we start by recalling how this test
works.

In what follows we extend the domain of our predicate $P: \{-1,1\}^k
\rightarrow \{0,1\}$ to $[-1,1]^k$ multi-linearly.  Thus, we
have $P: [-1,1]^k \rightarrow [0,1]$.

To prove hardness for Max-$P$, one analyzes the performance of the
dictatorship test in Figure~\ref{alg:Dictatorship test}, which uses a distribution
$\mu$ over $\{-1,1\}^k$ that we assume is supported on $P^{-1}(1)$ and
satisfies condition (1) of Theorem~\ref{thm:invariance}, which is the
case in both Theorems~\ref{thm:no neg hard} and
\ref{thm:resist_general}.

\begin{alg}{Dictatorship test}
  \alginput{A function $f: \{-1,1\}^L \rightarrow [-1,1]$}
  \algoutput{Accept/Reject}
\item Let $\mu_\epsilon = (1-\epsilon)\mu + \epsilon U_b$, where $U_b$ denotes the
product distribution over $\{-1,1\}^k$ where each bit has bias $b$.
\item Pick $L$ independent samples from $\mu_\epsilon$, giving $k$
  vectors $\vec x_1, \ldots, \vec x_k \in \{-1,1\}^L$.
\item Accept with probability $P(f(\vec x_1), \ldots, f(\vec x_k))$.
\end{alg}

The completeness property of the test is easy to establish (and only
depends on $\mu$ being supported on strings accepted by $P$).

\begin{lemma}
  If $f$ is a dictatorship function then $A$ accepts
  with probability $\ge 1-\epsilon$.
\end{lemma}

For the soundness analysis, the arguments are going to be slightly
different for the two Theorems~\ref{thm:no neg hard} and
\ref{thm:resist_general}.  It is convenient to view $f$ in its multilinear
form as described in the previous section.  Thus, instead of looking
at $f(\vec x_1), \ldots, f(\vec x_k)$ we look at $F(X_1), \ldots,
F(X_k)$.  In both cases, the goal is to prove that there are $d$ and
$\tau$ such that if $F$ is $(d,\tau)$-quasirandom then the expectation
of $Q(F(X_1), \ldots, F(X_k))$ is small (at most $E_Q + \epsilon$ for
Theorem~\ref{thm:resist_general} and at most $E_Q^+ + \epsilon$ for
Theorem~\ref{thm:no neg hard}).

In general, it is also convenient to apply the additional guarantee that
$F$ is balanced (i.e., satisfying $\E[F(X)] = 0$).  This can be achieved
by the well-known trick of folding, and is precisely what causes the
resulting Max-$P$ instances to have negated literals.  In other words,
when we prove Theorem~\ref{thm:no neg hard} on the hardness of
Max-$P^+$, we are not going to be able to assume this.

\paragraph{Theorem~\ref{thm:resist_general}: Relaxed Approximation Resistance Conditions}

The precise soundness property for Theorem~\ref{thm:resist_general} is
as follows.

\begin{lemma}
  Suppose $\mu$ is a distribution as in the statement of
  Theorem~\ref{thm:resist_general} and that the function $F$ is
  folded.  Then for every $\epsilon > 0$ there are $d, \tau$ such
  that whenever $F$ is $(d, \tau)$-quasirandom then 
  \[
  \E[Q(F(X_1), \ldots, F(X_k))] \le E_Q + \epsilon.
  \]
\end{lemma}

Note that in this case, the distribution $\mu_\epsilon$ is unbiased, in which
case the distribution of each column of $X$ is simply the distribution
$\mu_\epsilon$ itself.
\begin{proof} 
  We write $Q(x) = \sum_{S \subseteq [k]} \hat{Q}(S) \prod_{i \in S} x_i$,
  where $\hat{Q}(S)$ are the Fourier coefficients of $Q$ with respect to the
  uniform distribution, and $\hat{Q}(\emptyset) = E_Q$.

  We analyze the expectation of $Q$ term by term.  Fix a set
  $\emptyset \ne S \subseteq [k]$ and let us analyze $\E[\prod_{i \in
      S} F(X_i)]$.  Let $d,\tau$ be the values given by
  Theorem~\ref{thm:invariance}, when applied with $\epsilon$ chosen as
  $\epsilon/2^k$ and the $\alpha$ given by the distribution
  $\mu_\epsilon$ (note that this distribution satisfies the conditions
  of Theorem~\ref{thm:invariance}).    There are two cases.

  \begin{description}
  \item[Case 1: $|S| = 2$] Let $S = \{i,j\}$.  The conditions on $\mu$
    guarantee that $\mu$ is $\{i,j\}$-negative with respect to $Q$,
    i.e., for any column $a$ we have $\E[X^{a}_i] = \E[X^a_j] = 0$ and
    $\hat{Q}(S)\E[X^a_iX^a_j] \le 0$.  Let $\rho = \E[X^a_i X^a_j]$
    (as each column $a$ is identically distributed this value does not
    depend on $a$).  Then we have
    $$\hat{Q}(S)\E[F(X_i)F(X_j)] = \hat{Q}(S) \Stab_\rho(F)$$
    where $\Stab_\rho(F)$ is the noise stability of $F$ at $\rho$.
    Moreover, since the function $F$ is folded it is an odd function,
    which in particular implies that $\Stab_\rho(F)$ is odd as well so
    that $\sgn(\Stab_\rho(F)) = \sgn(\rho)$.  Since $\hat{Q}(S) \rho \le
    0$ it follows that $\hat{Q}(S) \Stab_\rho(F) \le 0$ as well, so $S$
    can not even give a positive contribution to the acceptance
    probability.

   \item[Case 2: $|S| \ne 2$, $\hat{Q}(S) \ne 0$] This is the more
    interesting case.  The conditions on $\mu$ guarantee that $\mu$
    covers $S$, i.e., there is an $i^* \in S$ such that $\E[X^a_{i^*}] = 0$
    and $\E[X^a_{i^*}X^a_j] = 0$ for all $j \in S \setminus \{i^*\}$.  By
    Theorem~\ref{thm:invariance}, we know that if $F$ is $(d,
    \tau)$-quasirandom we have
    $$
    \left|\E\left[\prod_{i \in S} F(X_i)\right] - \E\left[\prod_{i \in S} F(Y_i)\right]\right| \le \epsilon/2^k,
    $$ where $Y$ is a jointly Gaussian matrix with the same first and
    second moments as $X$.  But then, for every column $a$, the
    conditions on the second moments imply that $Y^a_{i^*}$ is a
    standard Gaussian completely independent from all other entries of
    $Y$.  This implies that $$\E\left[\prod_{i \in S} F(Y_{i^*})\right] =
    \E\left[F(Y_{i^*})\right] \cdot \E\left[\prod_{i \in S\setminus \{i^*\}} F(Y_i)\right] =
    0,$$ where the second equality is by the assumption that $F$ is
    folded.  This implies that $S$ has at most a negligible
    contribution of $\epsilon/2^k$ to the acceptance probability of
    the test.
  \end{description}
\end{proof}

\paragraph{Theorem~\ref{thm:no neg hard}: No Negations}

As mentioned earlier, in the case when negated literals are not
allowed, we can no longer assume that $F$ is folded.  Furthermore, the
distribution $\mu$ over $\{-1,1\}^k$ used is only assumed to be
pairwise uniformly correlated.  The precise soundness is as follows.

\begin{lemma}
  Suppose $\mu$ is a uniformly positively correlated distribution.
  Then for every $Q: [-1,1]^k \rightarrow [-1,1]$ and $\epsilon > 0$ there are $d, \tau$ such that whenever
  $F$ is $(d, \tau)$-quasirandom then 
  \[
  \E[Q(F(X_1), \ldots, Q(X_k))] \le E_Q^+ + \epsilon.
  \]
\end{lemma}

\begin{proof}
  Similarly to the previous lemma, we are going to take $d, \tau$ to
  be the values given by Theorem~\ref{thm:invariance} with $\epsilon$
  chosen as $\frac{\epsilon}{2 \cdot 2^k}$.

  Note that since $\mu_\epsilon$ is a combination of $\mu$ and $U_b$,
  both being uniformly positively correlated, $\mu_\epsilon$ is also
  uniformly positively correlated.

  Let $b = \E_{x \sim \mu_\epsilon}[x_i]$ and $\rho = \E_{x \sim \mu_\epsilon}[x_ix_j]
  \ge b^2$ be the bias and correlation of $\mu_\epsilon$, respectively.
  Define a new distribution $\eta$ over $\{-1,1\}^k$ as
  $$
  \eta =  c U_{\sqrt{\rho}} + (1-c) U_{-\sqrt{\rho}},
  $$
  where $c = \frac{b+\sqrt{\rho}}{2\sqrt{\rho}} \in [0,1]$ (recall that
  $U_{\sqrt{\rho}}$ denotes the product distribution where all biases
  are $\sqrt{\rho}$).

  Then $\eta$ has the same first and second moments as $\mu_\epsilon$
  and therefore, writing $Z$ for the corresponding matrix from $\eta$, we can apply Theorem~\ref{thm:invariance} twice and see that for every $S \subseteq [k]$
  $$
  \left|\E\left[\prod_{i \in S}F(X_i)\right] - \E\left[\prod_{i \in S} F(Z_i)\right]\right| \le \epsilon/2^k,
  $$
  implying
  $$
  \left|\E[ Q(F(X_1), \ldots, F(X_k))] - \E[Q(F(Z_1), \ldots, F(Z_k))]\right| \le \epsilon.
  $$ 

  Now, the column $Z^1$ of $Z$ can be written as a convex combination
  of two product distributions $R^+$ and $R^-$ over $\R^k$ (resulting
  from applying the character $\chi$ to $U_{\sqrt{\rho}}$ and
  $U_{-\sqrt{\rho}}$, respectively).  By linearity of expectation, we
  can replace $Z^1$ with one of $R^+$ and $R^-$ without decreasing the
  expectation of $Q(F(\cdot), \ldots, F(\cdot))$.  Repeating this for all columns, we end up with
  a random matrix $W$, each column of which is either distributed like
  $R^+$ or like $R^-$, and satisfying
  $$
  \E[Q(F(W_1), \ldots, F(W_k))] \ge \E[Q(F(Z_1), \ldots, F(Z_k))].
  $$ 
  But now since each column of $W$ is distributed according to a
  product distribution (with identical marginals), the rows of $W$ are
  independent and identically distributed, implying that 
  $$\E[Q(F(W_1), \ldots, F(W_k))] \le E_Q^+.$$
  Combining all our inequalities, we end up with
  $$
  \E[Q(F(X_1), \ldots, F(X_k))] \le E_Q^+ + \epsilon,
  $$
  as desired.
\end{proof}

\section{Concluding Remarks}\label{sec:concl}

We have introduced a notion of (computational) uselessness of
constraint satisfaction problems, and showed that, assuming the unique
games conjecture, this notion admits a very clean and nice
characterization.  This is in contrast to the related and more
well-studied notion of approximation resistance, where the indications
are that a characterization, if there is a reasonable one, should be
more complicated.

Our inability to obtain any non-trivial NP-hardness results
for positive uselessness, instead of Unique Games-based hardness
is frustrating.
While \cite{jhacm} proves odd parity of four variables to be
positively approximation resistant, obtaining positive uselessness by
the same method appears challenging.

Another direction of future research is understanding
uselessness in the completely satisfiable case.
%A third question is the one mentioned in Section~\ref{sec:completely
%  approx} about whether there exists a predicate $P$ that is
%computationally useful for all predicates or functions for whi

We have focused on CSPs defined by a single predicate $P$ (with or
without negated literals).  It would be interesting to generalize the
notion of usefulness to a general CSP (defined by a family of
predicates).  Indeed, it is not even clear what the correct definition
is in this setting, and we leave this as a potential avenue for future
work.  Another possible direction is to consider an analogous notion
for the decision version of a CSP rather than the optimization
version.

\noindent
{\bf Acknowledgement.}  We are grateful to a number of
anonymous referees for useful comments on the presentation
of this paper.

\bibliographystyle{alpha}
\bibliography{cspbib}

\end{document}